\documentclass[preprint]{elsarticle}
\usepackage[english]{babel}
\usepackage{verbatim}
\usepackage{alltt}
\usepackage{graphicx}
\usepackage{multirow}
\usepackage{calc}
\usepackage{amssymb}
\usepackage{amstext}
\usepackage{amsmath}

\usepackage[centering,includeheadfoot,margin=3.4cm]{geometry}

\makeatletter
\g@addto@macro\@verbatim\small
\makeatother

\begin{document}

\title{Scanning and Parsing Languages with Ambiguities and Constraints: The Lamb and Fence Algorithms}
\author{Luis~Quesada, Fernando~Berzal, and Francisco J.~Cortijo\\
  Department of Computer Science and Artificial Intelligence, \\
  CITIC, University of Granada, \\ 
  Granada 18071, Spain \\
  \textit{\{lquesada, fberzal, cb\}@decsai.ugr.es}
  }

\begin{abstract}

Traditional language processing tools constrain language designers to specific kinds of grammars.
In contrast, model-based language processing tools decouple language design from language processing.
These tools allow the occurrence of lexical and syntactic ambiguities in language specifications and the declarative specification of constraints for resolving them.
As a result, these techniques require scanners and parsers able to parse context-free grammars, handle ambiguities, and enforce constraints for disambiguation.
In this paper, we present Lamb and Fence.
Lamb is a scanning algorithm that supports ambiguous token definitions and the specification of custom pattern matchers and constraints.
Fence is a chart parsing algorithm that supports ambiguous context-free grammars and the definition of constraints on associativity, composition, and precedence, as well as custom constraints.
Lamb and Fence, in conjunction, enable the implementation of the ModelCC model-based language processing tool.


\end{abstract}

\maketitle

\section{Introduction}
\noindent
Traditional language processing tools \cite{Aho1972} require the textual specification of the language grammar in order to produce scanners and parsers for the desired language.

In contrast, model-based language processing tools \cite{Kleppe2007} automatically generate a grammar for the desired language (as well as the corresponding scanners and parsers for the language) given a conceptual model of the language.

Model-based language processing tools encourage model reuse and combination in order to reduce developer workload.
The combination of separate parts of different models could yield language ambiguities.
For example, a model considers, among other language elements, integers and points, and another model considers, among other elements, decimals; when combining parts of both models into another model, the ``5.2'' string could be interpreted either as a sequence of \emph{integer point integer} tokens, or as a single \emph{decimal} token.

Even the specification of a single model that considers both \emph{integer}, \emph{point}, and \emph{decimal} tokens is more elegant when ambiguities are allowed, since the ambiguity does not have to be minded or worked around.
It should be noted that the typical workaround, that is, specifying that the \emph{decimal} symbol is composition of two \emph{integer} symbols separated by a \emph{point} symbol, is semantically incorrect, since a decimal does not consist of two integers.

Lexical ambiguities occur when an input string simultaneously corresponds to several token sequences \cite{Nawrocki1991}, in which tokens may overlap.
Syntactic ambiguities occur when a token sequence can be parsed into several parse trees \cite{Aho1975}.

Therefore, model-based language processing tools require ambiguity support in both the scanner and the parser (or just in the parser, if we dispense with the lexical analyzer in scannerless parsing).

The language model generated by a model-based parser generator may contain epsilon productions (e.g. $A := \epsilon$), recursive productions (e.g. $A := A b$), grammar cycles (e.g. $A := c$, $A := B$, and $B := A$), constraints on associativities, compositions, and precedences, as well as custom constraints.
Therefore, a parsing algorithm that supports these constructions and constraints is required.

In this paper, we describe such a parsing algorithm, Fence, and its accompanying scanning algorithm, Lamb, in detail.

Lamb \cite{Quesada2011a} is a scanning algorithm that captures all the possible sequences of tokens within a given input string and produces a lexical analysis graph that describes them all.

Fence \cite{Quesada2012f} is bottom-up chart parsing algorithm that accepts lexical analysis graphs as input, performs an efficient syntactic analysis of them by taking constraints into account, and produces parse graphs that represent all the valid parse trees.

The Fence parsing algorithm complements the Lamb scanning algorithm by performing an implicit context-sensitive analysis of the lexical analysis graph provided by Lamb, as the parsing process determines which token sequences are valid in the parse graph and discards incorrect token sequences.

In addition, Fence supports every possible construction in a context-free language grammar, including epsilon productions and recursive productions.

Both Lamb and Fence are part of the ModelCC model-based parser generator \cite{Quesada2011c,Quesada2011f,Quesada2012k}, a powerful tool that demonstrates the feasibility of the model-driven approach we advocate. The ModelCC suite is open-source and it can be downloaded from http://www.modelcc.org.

The organization of this paper is as follows.
Section \ref{sec:background} describes formal grammars, surveys scanning and parsing algorithms, reviews scanner and parser generators, and introduces the model-driven language specification approach. 
Section \ref{sec:lamb} presents the Lamb scanning algorithm, whereas Section \ref{sec:fence} presents the Fence parsing algorithm.
Finally, Section \ref{sec:conclusionsfuturework} summarizes our conclusions and provides some pointers for further research.

\section{Background} \label{sec:background}
\noindent
In this section, we introduce context-free grammars (Subsection \ref{subsec:formalgrammars}), describe the typical architecture of language processor front-ends (Subsection \ref{subsec:architecture}), survey key parsing algorithms (Subsection \ref{subsec:scanningparsing}), review existing scanner and parser generators (Subsection \ref{subsec:generators}), and introduce model-based language specification techniques and the ModelCC model-based parser generator (Subsection \ref{subsec:modelbased}).

\subsection{Context-Free Grammars} \label{subsec:formalgrammars}

Context-free grammars are used to specify the syntax of context-free languages \cite{Aho2006}.
Using a set of rules, a context-free grammar describes how to form strings from the language alphabet that are valid according to the language syntax.
A context-free grammar $G$ is formally defined \cite{Chomsky1956} as the tuple $(N,\Sigma,P,S)$, where:

\begin{itemize}
\item $N$ is the finite set of nonterminal symbols of the language, sometimes called syntactic variables, none of which appear in the language strings.
\item $\Sigma$ is the finite set of terminal symbols of the language, also called tokens, which constitute the language alphabet (i.e. they appear in the language strings).
Therefore, $\Sigma$ is disjoint from $N$.
\item $P$ is a finite set of productions, each one of the form $N \rightarrow (\Sigma \cup N)^{*}$, where $*$
is the Kleene star operator, $\cup$ denotes set union, the part before the arrow is called the LHS (left-hand side) of the production, and the part
after the arrow is called the RHS (right-hand side) of the production.
\item $S$ is a distinguished nonterminal symbol, $S \in N$: the grammar start symbol.
\end{itemize}

For convenience, when several productions share their left-hand side, they can be grouped into a single production containing their the shared left-hand side and all their different right-hand sides separated by $|$.

A context-free grammar is said to be ambiguous if there exists at least one string that can be generated by the grammar in more than one way.
In fact, some context-free languages are inherently ambiguous (i.e. all context-free grammars generating them are ambiguous).

\subsection{The Architecture of Language Processors} \label{subsec:architecture}

The architecture of a language-processing system decomposes language processing into several steps which are typically grouped into two phases: analysis and synthesis.
The analysis phase, which is responsibility of the language processor front end, starts by breaking up its input into its constituent pieces (lexical analysis or scanning) and imposing a grammatical structure upon them (syntax analysis or parsing).
The language processor back end will later synthesize the desired target from the results provided by the front end.

A lexical analyzer, also called lexer or scanner, processes an input string conforming to a language specification and produces the tokens found within it.
Lexical ambiguities occur when a given input string simultaneously corresponds to several token sequences \cite{Nawrocki1991}, in which tokens may overlap.

A parser, also called syntactic analyzer, processes sequences of input tokens and determines their grammatical structure with respect to the given language grammar, usually in the form of parse trees.
In the absence of lexical ambiguities, the parser input consists of a stream of tokens, whereas it will be a directed acyclic graph of tokens when lexical ambiguities are present.
Syntactic ambiguities occur when a given set of tokens simultaneously corresponds to several valid parse trees \cite{Aho1975}.


\subsection{Scanning and Parsing Algorithms} \label{subsec:scanningparsing}

Scanning and parsing algorithms are characterized by the expression power of the languages to which they can be applied, their support for ambiguities or lack thereof, and the constraints they impose on language specifications.

Traditional lexical analyzers are based on a finite-state machine that is built from a set of regular expressions \cite{McNaughton1960}, each of which describes a token type. The efficiency of regular expression lexical analyzers is $O(n)$, being $n$ the input string length.

Efficient parsers for certain classes of context-free grammars exist. These include top-down LL parsers, which construct a leftmost derivation
of the input sentence, and bottom-up LR parsers, which construct a rightmost derivation of the input sentence.

LL grammars were formally introduced in \cite{Lewis1968}, albeit LL(k) parsers predate their name \cite{Oettinger1961}. An LL parser is called
an LL(k) parser if it uses $k$ lookahead tokens when parsing a sentence, while it is an LL(*) parser if it is not restricted to a finite set of $k$ lookahead tokens and it can make parsing decisions by recognizing whether the following tokens belong to a regular language
\cite{Jarzabek1975,Nijholt1976}, or by using syntactic or semantic predicates \cite{Parr1995}. While LL(k) parsers are always linear, LL(*) ranges from $O(n)$ to $O(n^2)$.

LR parsers were introduced by Knuth \cite{Knuth1965}. DeRemer later developed the LALR \cite{DeRemer1969,DeRemer1982} and SLR
\cite{DeRemer1971} parsers that are in use today. When parsing theory was originally developed, machine resources were scarce, and so parser
efficiency was the paramount concern \cite{Parr2011}. Hence all the aforementioned parsing algorithms parse in linear time (i.e. their
efficiency is $O(n)$, being $n$ the input string length) and they do not support syntactic ambiguities.

Efficient LR and LL parsers for certain classes of ambiguous grammars are also possible by using simple disambiguating rules \cite{Aho1975,Earley1975}.

A general-purpose dynamic programming algorithm for parsing context-free grammars was independently developed by Cocke \cite{Cocke1970}, Younger
\cite{Younger1967}, and Kasami \cite{Kasami1965}: the CYK parser. This general-purpose algorithm is $O(n^3)$ for ambiguous and unambiguous
context-free grammars. The Earley parser \cite{Earley1970} is another general-purpose dynamic programming algorithm for parsing context-free
grammars that executes in cubic time ($O(n^3)$) in the general case, quadratic time ($O(n^2)$) for unambiguous grammars, and linear time
($O(n)$) for almost all LR(k) grammars.

Free from the requirement to develop efficient linear-time parsing algorithms, researchers have developed many powerful nondeterministic parsing
strategies following both the top-down approach (LL parsers) and the bottom-up approach (LR parsers).

Following the top-down approach, Packrat parsers \cite{Ford2002packrat} and their associated Parsing Expression Grammars (PEGs)
\cite{Ford2004peg} preclude only the use of left-recursive grammar rules.
Even though they use backtracking, packrat parsers are linear rather than exponential because they memoize partial results, they attempt the alternative productions in the specified order, and they accept only the first one that matches an input position. In fact, LL(*) is an optimization of packrat parsing \cite{Parr2011}.

Following the bottom-up approach, Generalized LR (GLR) is an extension of LR parsers that handles nondeterministic and ambiguous grammars.
GLR forks new subparsers to pursue all possible actions emanating from nondeterministic LR states, terminating any subparsers that lead to invalid parses.
The result is, again, a parse forest with all possible interpretations of the input.
GLR parsers perform in linear to cubic time, depending on how closely the grammar conforms to the underlying LR strategy.
The time required to run the algorithm is proportional to the degree of nondeterminism in the grammar.
Bernard Lang is typically credited with the original GLR idea \cite{Lang1974}. Later, Tomita used the algorithm for natural language processing
\cite{Tomita1985}. Tomita's Universal parser \cite{Tomita1987}, however, failed for grammars with epsilon rules (i.e. productions with an
empty right-hand side). Several extensions have been proposed that support epsilon rules \cite{Farshi1991,Rekers1992,Ishii1994,McPeak2004}.

\subsection{Lexer and Parser Generators} \label{subsec:generators}

Lexer and parser generators are tools that take a language specification as input and produce a lexical analyzer or parser as output. They can be
characterized by their input syntax, their ability to specify semantic actions, and the parsing algorithms the resulting parsers implement.

Lex \cite{lex} and yacc \cite{yacc} are commonly used in conjunction \cite{Levine1992}. They are the default lexical analyzer generator and parser generator, respectively, in many Unix environments and standard compiler textbooks often use them as examples, e.g. \cite{Aho2006}. Lex is the
prototypical regular-expression-based lexical analyzer generator, while yacc and its many derivatives generate LALR parsers.

JavaCC \cite{McManis1996} is a parser generator that creates LL(k) parsers, albeit it has been superseded by ANTLR \cite{Parr1995}.
ANTLR is a parser generator that creates LL(*) parsers. ANTLR-generated parsers are linear in practice and greatly reduce
speculation, reducing the memoization overhead of pure packrat parsers.

The Rats! \cite{Grimm2006} packrat parser generator is a PEG-based tool that also optimizes memoization to improve its speed and reduce its
memory footprint. Like ANTLR, it does not accept left-recursive grammars. Unlike ANTLR, programmers do not have to deal with conflict messages,
since PEGs have no concept of a grammar conflict: they always choose the first possible interpretation, which can lead to unexpected behavior.

NLyacc \cite{Ishii1994} and Elkhound \cite{McPeak2004} are examples of GLR parser generators. Elkhound achieves yacc-like
parsing speeds when grammars are LALR(1).
Like PEG parsers, GLR parsers silently accept ambiguous grammars and programmers have to detect ambiguities dynamically
\cite{Parr2011}.

YAJco \cite{Poruban2009} is an interesting tool that accepts, as input, a set of Java classes with annotations that specify the prefixes,
suffixes, operators, tokens, parentheses, and optional elements common in typical programming languages. As output, YAJco generates a
BNF-like grammar specification for JavaCC \cite{McManis1996}. Since YAJco is built on top of a parser generator, the language designer
has to be careful when annotating his classes, as the implicit grammar he is defining has to comply with the constraints imposed by the
underlying LL(k) parser generator.

\subsection{Model-Based Language Specification} \label{subsec:modelbased}

In its most general sense, a model is anything used in any way to represent something else. In such sense, a grammar is a model of the language it defines.
In Software Engineering, data models are also common. Data models explicitly determine the structure of data. Roughly speaking, they describe the elements they represent and the relationships existing among them.
From a formal point of view, it should be noted that data models and grammar-based language specifications are not equivalent, even though both of them can be used to represent data structures. A data model can express relationships a grammar-based language specification cannot.
A data model does not need to comply with the constraints a grammar-based language specification has to comply with. Typically, describing a data model is generally easier than describing the corresponding grammar-based language specification.

In practice, when we want to build a complex data structure from the contents of a file, the implementation of the language processor needed to parse the file requires the software engineer to build a grammar-based language specification for the data as represented in the file and also to implement the conversion from the parse tree returned by the parser to the desired data structure, which is an instance of the data model that describes the data in the file.

Whenever the language specification has to be modified, the language designer has to manually propagate changes throughout the entire language processor tool chain, from the specification of the grammar defining the formal language (and its adaptation to specific parsing tools) to the corresponding data model. These updates are time-consuming, tedious, and error-prone. By making such changes labor-intensive, the traditional language processing approach hampers the maintainability and evolution of the language used to represent the data \cite{Kats2010}.

Moreover, it is not uncommon for different applications to use the same language. For example, the compiler, different code generators, and other tools within an IDE, such as the editor or the debugger, typically need to grapple with the full syntax of a programming language. Unfortunately, their maintenance typically requires keeping several copies of the same language specification synchronized.

The idea behind model-based language specification is that, starting from a single abstract syntax model (ASM) that represents the core concepts in a language, language designers can develop one or several concrete syntax models (CSMs). These CSMs can suit the specific needs of the desired textual or graphical representation for the language sentences. The ASM-CSM mapping can be performed, for instance, by annotating the abstract syntax model with the constraints needed to transform the elements in the abstract syntax into their concrete representation.

This way, the ASM representing the language can be modified as needed without having to worry about the language processor and the peculiarities of the chosen parsing technique, since the corresponding language processor will be automatically updated.

Finally, as the ASM is not bound to a particular parsing technique, evaluating alternative and/or complementary parsing techniques is possible without having to propagate their constraints into the language model. Therefore, by using an annotated ASM, model-based language specification completely decouples language specification from language processing, which can be performed using whichever parsing techniques are suitable for the formal language implicitly defined by the abstract model and its concrete mapping.

\begin{figure}[tb]
\begin{minipage}[tb]{\linewidth}
\centering
\includegraphics[scale=0.22]{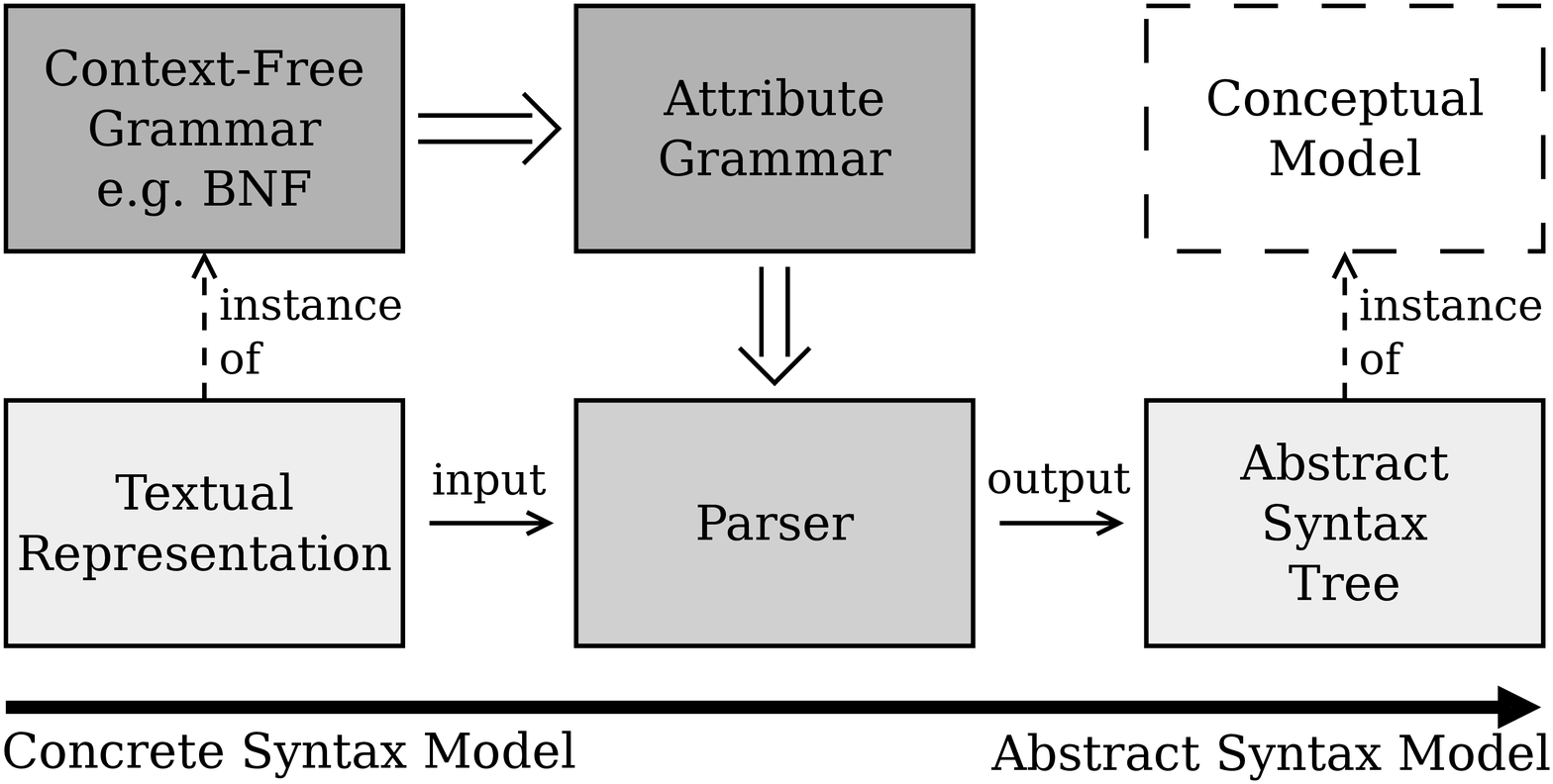}
\caption{Traditional language processing.} \label{fig:traditional}
\vspace{8mm}
\includegraphics[scale=0.22]{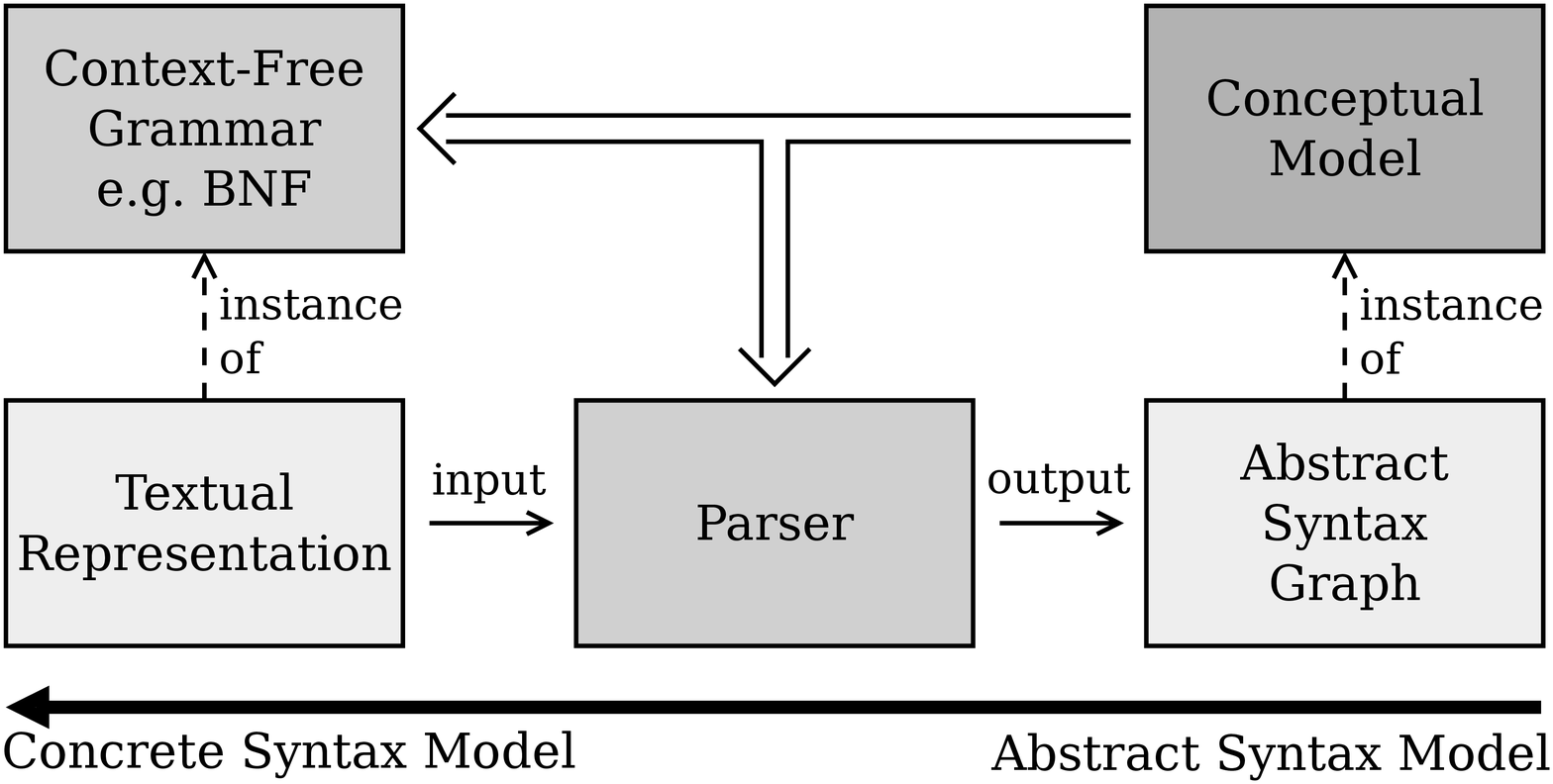}
\caption{Model-based language processing.} \label{fig:ModelCC}
\end{minipage}
\end{figure}

A diagram summarizing the traditional language design process is shown in Figure \ref{fig:traditional}, whereas the corresponding diagram for the model-based approach is shown in Figure \ref{fig:ModelCC}. It should be noted that ASMs represent non-tree structures whenever language elements can refer to other language elements, hence the use of the `abstract syntax graph' term.

The ASM is built on top of basic language elements, which can be viewed as the tokens in the model-driven specification of a language. Model-driven language processing tools such as ModelCC provide the necessary mechanisms to combine those basic elements into more complex language constructs, which correspond to the use of concatenation, selection, and repetition in the syntax-driven specification of languages.

When the ASM represents a tree-like structure, a model-based parser generator is equivalent to a traditional grammar-based parser generator in terms of expression power. When the ASM represents non-tree structures, reference resolution techniques can be employed to make model-based parser generators more powerful than grammar-based ones \cite{Quesada2012k}, as we will see in the next section.

For instance, in ModelCC, the constraints imposed over ASMs to define a particular ASM-CSM mapping can be declared as metadata annotations on the model itself. Now supported by all the major programming platforms, metadata annotations are often used in reflective programming and code generation \cite{Fowler2002}. Table \ref{fig:tablesummary} summarizes the set of constraints supported by ModelCC for establishing ASM-CSM mappings between ASMs and their concrete representation in textual CSMs.

\begin{table}[tb]
\begin{center}

\setlength{\tabcolsep}{5pt}
\begin{tabular}{ l  l  l } \hline

Constraints on... & Annotation & Function \\ \hline

\multirow{2}{*}{...patterns}
& @Pattern & Pattern matching definition of basic language elements. \\
& @Value & Field where the recognized input element will be stored. \\ \hline

\multirow{3}{*}{...delimiters}
& @Prefix & Element prefix(es). \\
& @Suffix & Element suffix(es). \\
& @Separator & Element separator(s) in lists of elements. \\ \hline

\multirow{3}{*}{...cardinality}
& @Optional & Optional elements.\\
& @Minimum & Minimum element multiplicity.\\
& @Maximum & Maximum element multiplicity.\\ \hline

\multirow{3}{40pt}{...evaluation order}
& @Associativity & Element associativity (e.g. left-to-right). \\
& @Composition & Eager or lazy composition for nested composites. \\
& @Priority & Element precedence level/relationships. \\ \hline

\multirow{2}{40pt}{...composition order}
& @Float & When element member position may vary. \\
& @FreeOrder & When all the element members positions may vary. \\ \hline

\multirow{2}{*}{...references}
& @ID & Identifier of a language element. \\
& @Reference & Reference to a language element. \\ \hline

\multirow{2}{40pt}{Custom constraints}
& \multirow{2}{*}{@Constraint} & \multirow{2}{*}{Custom user-defined constraint.} \\ \\ \hline
\end{tabular}

\end{center}
\caption{The metadata annotations supported by the ModelCC model-based parser generator.} \label{fig:tablesummary}
\end{table}

It should be noted that model-based language specification techniques allow lexical and syntactic ambiguities: each language element is defined as a separate and independent entity, even when their pattern specification or syntactic specification are in conflict. Therefore, model-based language specification techniques require a lexical analysis algorithm with ambiguity support and a parsing algorithm with lexical and syntactic ambiguity support.

When a language element in a model is composed of only optional members, epsilon productions (such as $E := \epsilon$) arise in the grammar.
Furthermore, the combination of selection and concatenation may produce grammar cycles (such as $A := c$, $A := B$, and $B := A$).
The formal grammars of languages specified using model-based techniques may also contain constraints on associativity, composition, and precedence, and custom constraints.
Therefore, a scanner and a parser that supports such requirements are needed.

\section{The Lamb scanning algorithm} \label{sec:lamb}
\noindent
Lamb \cite{Quesada2011a} is a scanning algorithm with ambiguity support.
Lamb also supports token precedence constraints and the hard-coded specification of custom pattern matchers and lexical or semantic constraints.

The power of the Lamb scanning algorithm resides in the usage of lexical analysis graphs to store and output the scanning results.
In these graphs, each token is linked to its preceding and following tokens and there may also be several starting tokens.
Each path in a lexical analysis graphs represents a different sequence of tokens that can be found within the input string.

For example, the token specification listed in Figure \ref{fig:tokens} is lexically-ambiguous, since any sequence of digits separated with points could be considered either \emph{Decimal} tokens or \emph{Integer Point Integer} token sequences.

\begin{figure}[tb]
\begin{verbatim}
   (-|\+)?[0-9]+              Integer
   (-|\+)?[0-9]+\.[0-9]+      Decimal
   \.                         Point
   \#                         Hash
   \$                         Dollar
\end{verbatim}
\caption{Regular expressions and token names in a lexically-ambiguous token specification.}
\label{fig:tokens}
\end{figure}

\begin{figure*}[tb]
\begin{minipage}[tb]{\linewidth}
\centering
\includegraphics[scale=0.23]{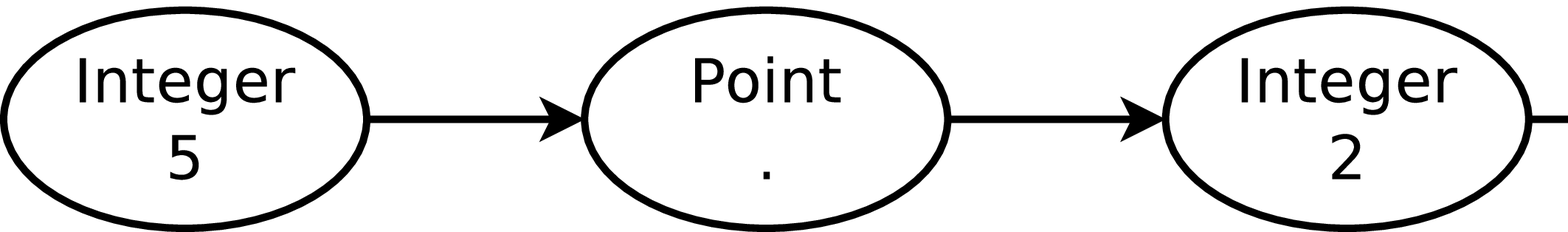}
\caption{Lexical analysis, as produced by a traditional scanning algorithm, when the \emph{Integer} token has a greater priority than the \emph{Decimal} token.}
\label{fig:e3}
\vspace{4mm}
\centering
\includegraphics[scale=0.23]{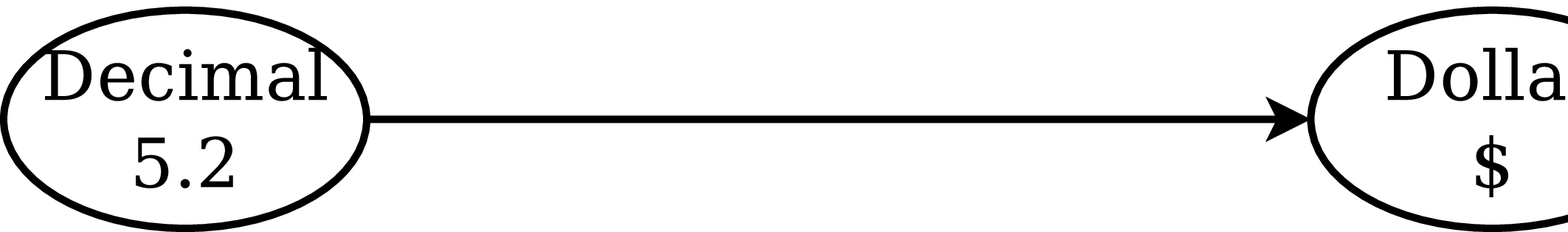}
\caption{Lexical analysis, as produced by a traditional scanning algorithm, when the \emph{Decimal} token has a greater priority than the \emph{Integer} token.}
\label{fig:e2}
\vspace{4mm}
\centering
\includegraphics[scale=0.23]{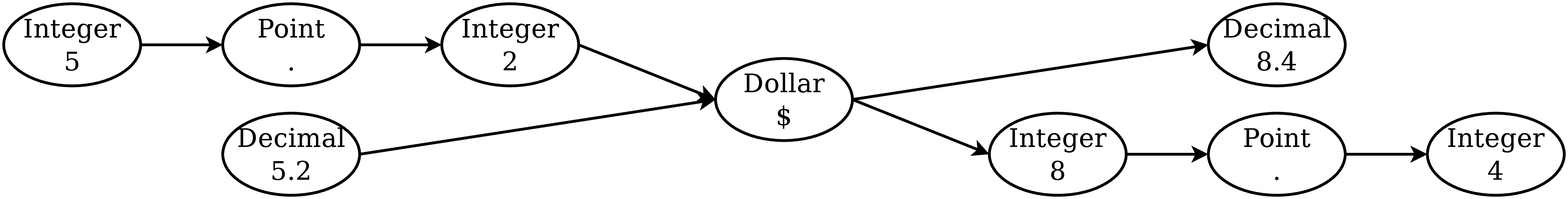}
\caption{Lexical analysis, as produced by Lamb, assuming that \emph{Decimal} and \emph{Integer} tokens do not have different priorities.}
\label{fig:e1}
\end{minipage}
\end{figure*}

Let us consider, for example, the ``5.2 \$ 8.4'' input string.
When using a traditional scanning algorithm, the developer can either assign the \emph{Integer} token a greater priority than the \emph{Decimal} token or vice versa.
The respective interpretations of the ``5.2 \$ 8.4'' string are shown in Figures \ref{fig:e3} and \ref{fig:e2}.

Unless specified by the language designer, Lamb does not impose token precedences and it is able to capture all the four possible token sequences in the form of a lexical analysis graph, as shown in Figure \ref{fig:e1}.

Lamb is able to recognize ambiguities that involve tokens of different types that overlap, in contrast to traditional approaches such as the Schr{\"o}dinger's token \cite{Aycock2001}, which cannot be applied to overlapping tokens as it only allows for strings to have a superposition of token types.
Indeed, in certain applications, even a string such as ``52'' could be interpreted as ambiguous: It could correspond either an \emph{Integer} or a sequence of two \emph{Integer} tokens.
Lamb allows choosing between a greedy pattern matching policy, in which given several tokens of the same type that start in the same position index, only the longest match will be considered; and an exploratory pattern matching policy, in which given several tokens of the same type that start in the same position index, all of them are considered.

In Section \ref{sec:fence}, we will illustrate the result of parsing this lexical analysis graph.

The Lamb scanning algorithm consists of two phases: the scanning phase, which recognizes all the possible tokens in the input string (Subsection \ref{subsec:scanning}); and the lexical analysis graph generation phase, which computes the sets of preceding and following tokens for each token (Subsection \ref{subsec:lag}).

\subsection{The Scanning Phase} \label{subsec:scanning}

The Lamb scanning phase matches a list of pattern matchers with the input string.
It can be configured to follow one of these two matching policies, depending on the language features:

\begin{itemize}
\item {\bf Greedy pattern matching policy.} Given several tokens of the same type that start in the same position index, only the longest match will be considered.
This policy avoids the generation of huge lexical analysis graphs, and can only be used when no tokens in the language can be broken up solely into two or more tokens of its same type with no additional delimitation.
If a token could be broken up into two or more tokens of its same type, the greedy policy would skip the subtokens.
For example, scanning the string ``4912'' for \emph{Integer} tokens by following the greedy pattern matching policy would perform a greedy matching and would yield the token ``4912''.
As the only found token starts at the first position in the input string and ends at the last position in the string, it is the only entity that could be delimiting any other tokens.
As there is nothing before the first position in the input string or after the last position in the input string, there are not any more tokens in it.
\item {\bf Exploratory pattern matching policy.} Given several tokens of the same type that start in the same position index, all of them are considered.
While this policy cannot render a lexical analysis incorrect, it allows the generation of a exponentially huge lexical analysis graphs, and is required when any tokens in the language can be broken up into two or more tokens of its same type with no additional delimitation.
For example, scanning the string ``4912'' for \emph{Integer} tokens by following the exploratory pattern matching policy would yield all the tokens ``4'', ``9'', ``49'', ``1'', ``91'', ``491'', ``2'', ``12'', ``912'', and ``4912''.
\end{itemize}

When performing the lexical analysis, Lamb keeps a forbidden token list and a state value for each position in the input string.
The forbidden token list stores constraints on the generation of different token types.
The state value determines whether a position index is a valid starting position for matchings (\emph{MATCH}) or not (\emph{SKIP}).
In the latter case, it should be noted that the character stored in that position can take part in matchings that started some positions before in the input string.

The greedy pattern matching initializes the state to \emph{MATCH} for the first position of the input string and to \emph{SKIP} for the rest of the positions of the input string.
For each position index of the input string, if its state is \emph{MATCH}, every pattern matcher corresponding to a token that is not in the forbidden token lists is applied in descending order of precedence.
The pattern matchers try to greedily match the single longest matching possible that starts in that position.
Whenever a matching is found, it adds the tokens it precedes to the forbidden token lists for all the positions of the matched text.
The state of the position index that follows the last symbol of the token is set to \emph{MATCH}, so that new tokens can start at that position.

The exploratory pattern matching assumes the state is \emph{MATCH} for all the input string positions.
Every pattern matcher is applied, in descending order of precedence, for each position in the input string.
The pattern matchers try to match every possible matching that starts in that position.
Whenever a matching is found, it adds the tokens it precedes to the forbidden token lists for all the positions of the matched text.

Either if the pattern matching policy is greedy or exploratory, this phase has a theoretical $O(ln^2)$ order of efficiency, being $n$ the input string length and $l$ the number of matchers in the scanner, even when the string can be tokenized in up to $l^n$ ways.

\subsection{The Lexical Analysis Graph Generation Phase} \label{subsec:lag}

The sets of preceding and following tokens of an $x$ token are defined in Equation \ref{eq:nextprev}, being $a,b,c$ tokens and $x_{start}$ and $x_{end}$ the starting and ending positions of the $x$ token in the input string.

\begin{equation}
\begin{split}
b \in &FOLLOWING(a) \wedge a \in PRECEDING(b)\textrm{ iif } \\
& (a_{end}<b_{start}) \wedge (\nexists c, c_{start}>a_{end} \wedge c_{end}<b_{start}) 
\end{split}
\label{eq:nextprev}
\end{equation}

Once these sets have been computed for every token, any token whose preceding set is empty is added to the start token set of the lexical analysis graph.

Either if the pattern matching policy is greedy or exploratory, this phase has a theoretical order of efficiency of $O(tk)$, being $t$ the number of tokens found and $k$ the maximum number of tokens that follow a token in the graph. As $t \leq n \cdot l$, the theoretical order of efficiency of this phase is $O(lkn)$.

The scanning and graph generation phases of greedy Lamb scanners have an overall $O(l(k+n)n)$ order of efficiency.

\section{The Fence parsing algorithm} \label{sec:fence}
\noindent
Fence is a bottom-up chart parsing algorithm with lexical and syntactic ambiguity support.
Fence also supports precedence, associativity, and composition constraints, as well as the hard-coded implementation of custom syntactic and semantic constraints.
Fence is able to parse languages described by grammars that contain epsilon productions, recursive productions, and grammar cycles.

The power of the Fence parser resides in the fact that constraints are enforced as soon as possible, thus partial parse graphs that do not comply with the constraints are promptly discarded.

\begin{figure*}[tb]
\begin{minipage}[tb]{\linewidth}
\centering
\begin{verbatim}
Product ::= Reference Price | Price Reference
Reference ::= [ Hash ] Integer Point Integer
Price ::= [ Dollar ] Decimal
\end{verbatim}
\caption{Context-sensitive syntactic rules needed to resolve the lexical ambiguities.}
\label{fig:srules}
\vspace{4mm}
\centering
\includegraphics[scale=0.23]{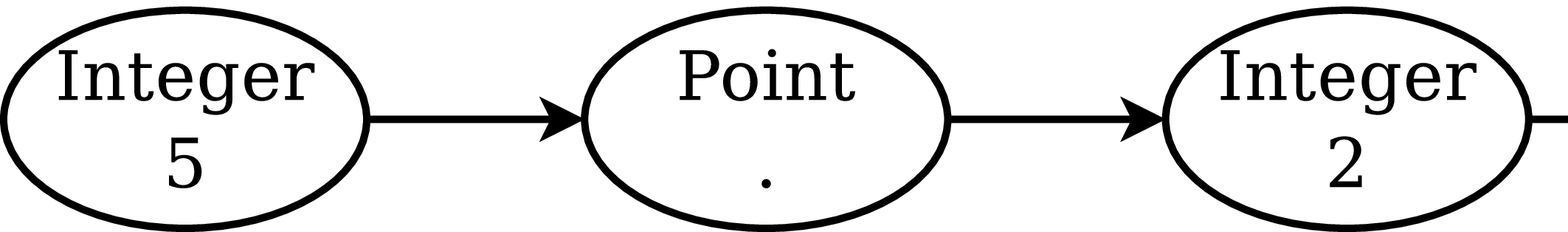}
\caption{Intended lexical analysis.}
\label{fig:e4}
\vspace{4mm}
\includegraphics[scale=0.23]{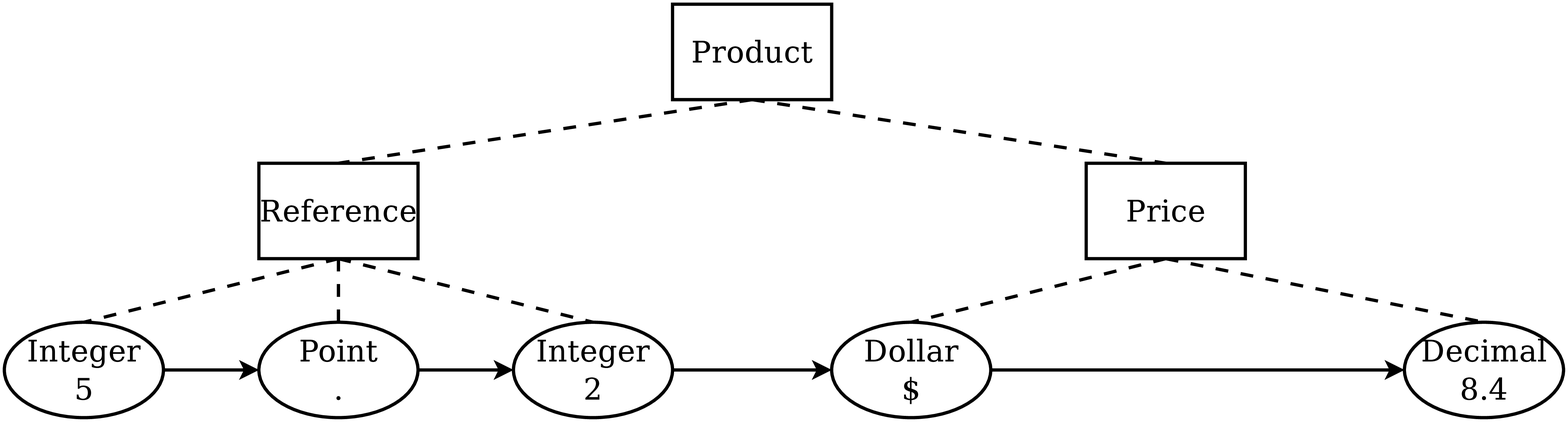}
\caption{The only valid syntax analysis tree produced by parsing the lexical analysis graph in Figure \ref{fig:e1} with the grammar in Figure \ref{fig:srules}.}
\label{fig:e5}
\end{minipage}
\end{figure*}

Continuing our example, the grammar productions shown in Figure \ref{fig:srules} complement the token specification described in the previous Section.
Given this grammar, the expected token sequence for the input string ``5.2 \$ 8.4'' is shown in Figure \ref{fig:e4}.

The language illustrates a scenario of lexical ambiguity sensitivity, as the tokens found in an input string depend on the context: a string such as ``5.2'' can be interpreted as a \emph{Decimal} token when forming part of a \emph{Price} symbol, or as an \emph{Integer} \emph{Point} \emph{Integer} token sequence when forming part of a \emph{Reference} symbol.
Since both the \emph{Dollar} and the \emph{Hash} components are optional in \emph{Price} and \emph{Reference} symbols, respectively, the disambiguation cannot be done in the scanner.
The appearance of either a \emph{Dollar} token or a \emph{Hash} token is nonetheless enough for identifying the \emph{Price} and \emph{Reference} symbols and reducing them to a valid \emph{Product} symbol.

Parsing the lexical analysis graph in Figure \ref{fig:e3} would produce a single valid parse tree, which, in turn, is based on the only lexical analysis that yields a valid parse tree given the grammar in the example.
The resulting parse tree is shown in Figure \ref{fig:e5}.

The Fence parsing algorithm consists of three phases: the extended lexical analysis graph construction phase, which prepares the lexical analysis graph to store parsing-related information (Subsection \ref{subsec:ela}); the chart parsing phase, which performs grammar production reductions (Subsection \ref{subsec:parsing}); and the constraint enforcement phase, which discards constructions not allowed by the constraints imposed on the language (Subsection \ref{subsec:constraints}).

\subsection{Extended Lexical Analysis Graph Construction Phase} \label{subsec:ela}

In order to efficiently perform the parsing process, Fence converts the input lexical analysis graph (LA graph) into an extended lexical analysis graph (ELA graph) that stores information about partially applied productions in data structures.

In the ELA graph, a dotted production of the form $N \rightarrow (\Sigma \cup N)^{*}.(\Sigma \cup N)^{*}$ indicates that the RHS symbols before the dot have already been matched with a substring of the input string.
A partially applied production (namely, handle) is a tuple $(dotted production,$ $[start,end])$, where $start$ and $end$ identify the substring of the input string that matched the dotted production RHS symbols before the dot.
Each handle can be used during the parsing process to match a rule RHS symbol with a node representing either a token or a nonterminal symbol (namely, SHIFT actions in LR-like parsers) or perform a reduction (namely, REDUCE actions in LR-like parsers).
A core is a set of handles.

\begin{figure*}[tb]
\centering
\includegraphics[scale=0.23]{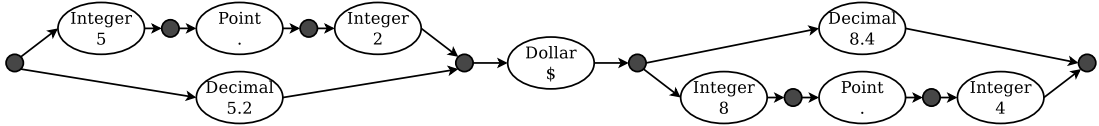}
\caption{Extended lexical analysis graph corresponding to the lexical analysis graph shown in Figure \ref{fig:e1}. Grayed nodes represent cores.}
\label{fig:e6}
\end{figure*}

In an ELA graph, tokens are not linked to their preceding and following tokens, but to their preceding and following cores.
Cores are, in turn, linked to their preceding and following token sets.
For example, the ELA graph corresponding to the LA graph in Figure \ref{fig:e1} is shown in Figure \ref{fig:e6}.

The conversion from the LA graph to the ELA graph is performed by completing the LA graph with cores.
A \emph{starting} core is linked to the tokens with an empty preceding token set.
A \emph{final} core is linked from the tokens with an empty following token set.
Finally, for each one of the other tokens in the LA graph, a preceding core is linked to it.
Links between tokens in the LA graph are converted into links from tokens to the cores preceding each token of their following token set in the ELA graph.

\subsection{Chart Parsing Phase} \label{subsec:parsing}

The Fence chart parsing phase processes the ELA graph and generates an implicit parse graph (I-graph).
Nodes in the I-graph are described as $(start,end,symbol)$ tuples, where $start$ and $end$ identify a substring in the input string, and $symbol$ identifies the production LHS.
Since I-graph nodes contain no information about their contents, both lexical and syntactic ambiguities are implicit (i.e. a GLR shared packed parse forest \cite{Tomita1985} without information on alternate subtrees).
The I-graph contains a set of starting nodes, each of which may represent several parse tree roots.
The parsing itself is performed by progressively applying productions and storing handles in cores.

The grammar productions with an empty RHS (i.e. epsilon productions) are removed from the grammar and their LHS symbol is stored in the \emph{epsilonSymbols} set.
This set allows these parse symbols being skipped when found in a production, as if a reduction using the epsilon production were applied.

The agenda is a stack of $(handle,node)$ pairs in which the node can match the symbol after the dot in the dotted rule of the handle. It is initially empty.
All the agenda entries ever generated are stored and checked in order to avoid the generation of duplicate entries.

The parser is initialized by generating a handle for each production and adding them to every core.

The parsing process consists of iteratively extracting entries consisting of handles and nodes from the agenda and matching the next symbol of the RHS of the handle production with the node.
The handles whose productions are successfully matched are added to the cores following the node and the agenda is updated with the entries that contain any of the newly generated handles.
In case all the symbols of a production RHS match a sequence of nodes, a new node is generated by reducing them.
The new node $start$ index is obtained from the handle, its $end$ position is obtained from the last matched node, and its $symbol$ is the LHS symbol of the reduced production.
When a newly generated node has only the \emph{starting} core in its preceding core set and the \emph{final} core in its following core set, and its $symbol$ corresponds to the initial symbol of the grammar, it is added to the parse graph starting node set, which means that that node represents a valid parse tree.

The result of the chart parsing phase is an I-graph, which the constraint enforcement phase accepts as input.

\subsection{Constraint Enforcement Phase} \label{subsec:constraints}

The Fence constraint enforcement phase processes the I-graph and generates an explicit parse graph (E-graph, or just parse graph) by expanding the implicit nodes and enforcing the constraints defined for the language.
Nodes in the E-graph that represent tokens are still defined as $(start,end,symbol)$ tuples.
Nodes in the E-graph that represent nonterminal symbols reference the list of nodes that matched the production used to generate those nodes (i.e. a structure similar to a GLR shared packed parse forest \cite{Tomita1985} in which only the nodes that comply with the constraints exist).
It should be noted that ambiguities, both lexical and syntactic, are explicit in the E-graph, as it represents several parse trees corresponding to all the possible interpretations of the input string.
The E-graph contains a set of starting nodes, each of which represents a parse tree root.
Constraint enforcement is performed by converting each implicit node into every possible explicit node sequence that can be derived from the implicit node and satisfies the specified constraints; that is, by expanding the each implicit node.

Only the nodes that conform valid parse trees are needed in the parse graph. In order to generate only these nodes, each one of the implicit nodes in the starting node set of the I-graph is recursively expanded using memoization.
Each possible resulting explicit node is the root of a parse tree in the E-graph.

The expansion of an implicit node is performed by finding every possible reduction of a sequence of explicit nodes that generates that node.
In order to do that, the implicit node is expanded by applying every possible production that could generate it and producing a set of explicit nodes.
Productions are applied by matching their RHS with explicit nodes.
Whenever an implicit node is found and needed in order to make the production application progress, it is expanded recursively.
It should be noted that this procedure is different from parsing itself in that the actual bounds of the reductions for every node are known.

A history set is stored for each node, in order to avoid expanding a node over a fragment of the input string as an indirect requirement of expanding the very same node.
For example, consider the grammar productions $A := c$, $A := B$, and $B := A$, the input string ``c'', and the $A$ parse tree root symbol that matches the only input character.
The $A$ symbol can be expanded using the $A := B$ production in order to generate a $B$ symbol that would match the only input character.
The $B$ symbol can be expanded using the $B := A$ production in order to generate another $A$ symbol that, again, matches the only input character.
The last $A$ symbol cannot be expanded again, as it occupies the very same input fragment than the $A$ symbol from two phases before.
This history avoids infinite loops in grammar cycles.

Once an explicit node is expanded, constraints are enforced on it and it is discarded if it does not comply with all of them.

Fence supports associativity constraints, selection precedence constraints, composition precedence constraints, and custom-designed constraints.

The fact that constraint checking is performed during graph expansion improves the parser performance, as the sooner constraints are applied, the larger number of interpretations are discarded.
For example, in the case of a binary expression with left-to-right associative operators, the string ``2+5+3+5+6+2+1+5+6+3'' can be expanded in $104857$ possible ways when not considering the associativity constraint, and in just one possible way when considering it.

\begin{figure*}[tb]
\centering
\includegraphics[scale=0.23]{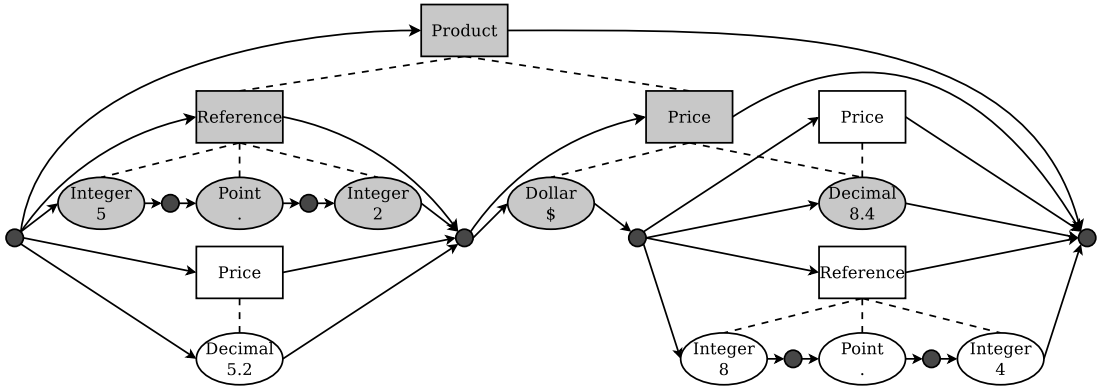}
\caption{Parse graph corresponding to the extended lexical analysis graph shown in Figure \ref{fig:e6}.
Squares represent nonterminal symbols found during the parsing process. Filled squares represent the intended parsing. Dotted lines represent the composition of the explicit parse graph.}
\label{fig:e7}
\end{figure*}

\begin{itemize}
\item {\bf Associativity constraints} allow the specification of the associative property for binary operators.
The application of a production is inhibited when any of the nodes that matches its RHS symbols has an associativity constraint and is followed (for left-to-right associativity constraints), preceded (for right-to-left associativity constraints), or either followed or preceded (for non-associative associativity constraints) by a node that was derived using the same production.

\item {\bf Selection precedence constraints} allow the resolution of syntactic ambiguities caused by different explicit nodes (i.e. interpretations) resulting from a single implicit node.
For example, a \emph{Statement} in a hypothetical imperative programming language could be either an \emph{OutputStatement} or a \emph{FunctionCall}. Both \emph{OutputStatement} and \emph{FunctionCall} can match the input string ``output(var);'', therefore \emph{OutputStatement} can be set to precede \emph{FunctionCall}, which will inhibit that string from being considered a function call.
The application of a production is inhibited when it is preceded by a different production and both of them match the same node sequence.

\item {\bf Composition precedence constraints} allow the resolution of syntactic ambiguities when a node derived using a production cannot be derived using another production.
For example, one of the productions \emph{ConditionalStatement ::= ``if'' Expression Sentence} and \emph{ConditionalStatement ::= ``if'' Expression Sentence ``else'' Sentence} can be set to precede the other one in order to resolve (in any of the two possible ways) the ambiguity of nested if-then-else statements in situations such as ``if expr1 if expr2 sent1 else sent2'', where ``else sent2'' could be assigned to either the inner or outer conditional sentence.
The application of a production is inhibited when it precedes any of the productions used to derive the nodes that matched its RHS symbols.

\item {\bf Custom constraints} allow the specification of any other constraints (e.g. semantic constraints).
In order to enforce custom-designed constraints, an evaluator can be assigned to any production. Whenever a node is generated, the evaluator of the production used to derive it is executed and determines whether the node satisfies the constraint or not. In the latter case, its generation is inhibited.
Custom-designed constraints provide a flexible framework which allows developers to design complex syntactic or semantic constraints (e.g. probabilistic constraints, corpus-based constraints) that effectively limit the possible interpretations of an input string and, as a side effect, improve the performance of the parser, as pruned partial interpretations are discarded as soon as they do not fulfill the constraints.
\end{itemize}

The result of the constraint enforcement phase is an E-graph or parse graph, such as the one shown in Figure \ref{fig:e7}.
The obtained parse graph, after stripping off all its ancillary structures, such as cores, equals the intended parse graph in Figure \ref{fig:e5}.

The Fence parsing algorithm executes, as Earley-type chart parsers, in cubic time ($O(n^3)$) in the general case, quadratic time ($O(n^2)$) for unambiguous grammars, and linear time ($O(n)$) for almost all LR(k) grammars.

The Fence constraint enforcement phase improves the performance of traditional techniques phases in practice, as all constraints are applied at the earliest possible time, thus discarding possibilities that would otherwise be processed later.

\section{Conclusions and Future Work} \label{sec:conclusionsfuturework}
\noindent

In this paper, we have presented the Lamb scanning algorithm and the Fence parsing algorithm.
Lamb supports ambiguous token definitions and the hard-coded specification of custom pattern matchers and constraints.
Fence supports ambiguous context-free grammars with epsilon productions, recursive productions, grammar cycles, and the definition of constraints on associativity, composition, and precedence, as well as user-defined syntactic and semantic constraints.

Lamb and Fence enable the implementation of model-based language specification tools.
Indeed, both Lamb and Fence are part of the ModelCC model-based parser generator, which is freely available at http://www.modelcc.org.

We used ModelCC (and, hence, Lamb and Fence) to implement the General Natural Language Parser \cite{Quesada2013b}.
In the near future, we intend to implement and evaluate natural language instances (e.g. English or Spanish) of the General Natural Language Parser.
We also intend to apply ModelCC to other application domains, such as data integration \cite{Tan2006} and model-driven software development tools \cite{Schmidt2006,mdsd-ideal}.

\bibliographystyle{abbrv}
\bibliography{doc}

\end{document}